\documentclass[fleqn,11pt]{wlscirep}
\usepackage[utf8]{inputenc}
\usepackage[T1]{fontenc}
\usepackage{cmbright}

\newcommand{\euclid}{{\it Euclid}}

\newcommand{\rom}{{\it Roman}}
\newcommand{\kepler}{{\it Kepler}}

\newcommand{\vis}{{\sc VIS}}
\newcommand{\nisp}{{\sc NISP}}

\newcommand{\murel}{\mu_{\mathrm{rel}}}

\title{Roman CCS White Paper\\
Magnifying NASA Roman GBTDS exoplanet science with coordinated observations by ESA Euclid}
\author[1,a,*]{The Euclid Exoplanet Science Working Group: Eamonn Kerins}
\author[2]{Etienne Bachelet}
\author[3,4]{Jean-Philippe Beaulieu}
\author[5,6]{Valerio Bozza}
\author[1,7]{Iain McDonald}
\author[8]{Matthew Penny}
\author[4]{Cl\'{e}ment Ranc}
\author[9]{Jason Rhodes}
\author[10,a]{Mar\'{i}a Rosa Zapatero Osorio}
\affil[1]{Department of Physics and Astronomy, The University of Manchester, Oxford Road, Manchester M13 9PL, UK.}
\affil[2]{IPAC, Mail Code 100-22, Caltech, 1200 E. California Blvd., Pasadena, CA 91125, USA}
\affil[3]{School of Physical Sciences, University of Tasmania, Private Bag 37 Hobart, Tasmania 7001, Australia}
\affil[4]{Institut d'Astrophysique de Paris, CNRS, Sorbonne Universit\'e, F-75014 Paris, France}
\affil[5]{Dipartimento di Fisica “E.R. Caianiello”, Università di Salerno, Via Giovanni Paolo II 132, I-84084 Fisciano, Italy}
\affil[6]{Istituto Nazionale di Fisica Nucleare, Sezione di Napoli, Via Cintia, 80126, Napoli, Italy}
\affil[7]{Open University, Walton Hall, Milton Keynes, MK7 6AA United Kingdom}
\affil[8]{Louisiana State University, 261-B Nicholson Hall, Tower Dr., Baton Rouge, LA 70803-4001, USA}
\affil[9]{Jet Propulsion Laboratory, California Institute of Technology, 4800 Oak Grove Drive, Pasadena, CA 91109, USA}
\affil[10]{Centro de Astrobiología, CSIC-INTA, Carretera de Ajalvir km 4, 28850 Torrejón de Ardoz, Madrid, Spain}
\affil[a]{Euclid Exoplanet Science Working Group Lead}
\affil[*]{Corresponding author: Eamonn.Kerins@manchester.ac.uk}

\begin{abstract}
The ESA \euclid{} mission is scheduled to launch on July 1st 2023. This White Paper discusses how \euclid{} observations of the Galactic Bulge Time Domain Survey (GBTDS) area could dramatically enhance the exoplanet science output of the Nancy Grace Roman Space Telescope (\rom{}). An early \euclid{} pre-imaging survey of the \rom{} GBTDS fields, conducted soon after launch, can improve proper motion determinations for \rom{} exoplanet microlenses that can yield a factor of up to $\sim 5$ improvement in exoplanet mass measurements. An extended \euclid{} mission would also enable the possibility of sustained simultaneous observations of the GBTDS by \euclid{} and \rom{} that would achieve large gains in several areas of \rom{} exoplanet science, including science that is impossible to achieve with \rom{} alone. These include: a comprehensive demographic survey for free-floating planets that includes precision mass measurements to establish the true nature of individual candidates; detection, confirmation and mass measurements of exomoons; direct exoplanet mass measurements through parallax and finite source size effects for a large sample of bound exoplanets detected jointly by \euclid{} and \rom{}; enhanced false-positive discrimination for the large samples of transiting planets that \rom{} will detect. Our main recommendation to NASA and ESA is to initiate a Joint Study Group as early as possible that can examine how both missions could best conduct a coordinated campaign. We also encourage flexibility in the GBTDS scheduling.
\end{abstract}
\begin{document}

\flushbottom
\maketitle
%
%
\thispagestyle{empty}

\noindent {\bf SCIENTIFIC CATEGORIES:} Exoplanets and exoplanet formation; Stellar populations and the interstellar medium
\\


\section{Introduction} \label{sec:intro}

The Galactic Bulge Time Domain Survey (GBTDS) is one of three Core Community Surveys that will be undertaken by the NASA Nancy Grace Roman Space Telescope (\rom{} -- \citealt{2015arXiv150303757S}). The principal driver for the GBTDS is to complete the statistical exoplanet census started by \kepler{} by detecting large numbers of cool planets down to Earth mass, and determining their occurrence and demography \citep{2019ApJS..241....3P}. Microlensing is the only available technique able to access this regime and, if conducted from space, such surveys can provide exoplanet yields comparable to \kepler{}. As a by-product of the survey design, \rom{} will also find some 60,000-200,000 distant transiting planets \citep{2023arXiv230516204W}. 

The purpose of this White Paper is to highlight a huge opportunity to extend significantly \rom{} exoplanet capabilities through coordinated observations with the ESA \euclid{} mission \citep{2011arXiv1110.3193L}. \euclid{} is scheduled to launch from Cape Canaveral at 15:42~GMT on July~1st 2023 (two weeks after submission of this White Paper), some 3.5 years before the anticipated launch of \rom{}. Working within the constraints set by the \euclid{} primary cosmology mission, we  advocate two types of \euclid{} observations that could assist the \rom{} GBTDS. First, in the coming months, \euclid{} will have an opportunity to observe the likely \rom{} GBTDS fields to provide a longer baseline over which to measure the motion of  planetary hosts of microlensing events detected by \rom{}. This will greatly improve many planet mass measurements, and may enable mass measurements for some events for which the \rom{} time baseline will be too short \citep{2022A&A...664A.136B}. Second, towards the end of the \euclid{} main mission, and into an extended mission, and in the first phases of the \rom{} mission, \euclid{} and \rom{} can simultaneously observe the GBTDS fields. These joint observations will bring huge improvements to \rom{} exoplanet measurements, and its capability to detect and confirm exomoons. It will also enable precision studies of free-floating planets that are impossible to achieve with \rom{} alone.

We discuss in this White Paper the science that can be achieved but also some potential technical and logistical obstacles. {\em Our principal recommendation is to encourage NASA and ESA to work together as early as possible to explore these opportunities through a formal Joint Study Group.}

This White Paper is organized as follows. In Section~\ref{sec:telescope} we overview the \euclid{} telescope and instrumentation. In Section~\ref{sec:micro} we briefly review microlensing theory to outline how direct planet mass measurements are obtained. Section~\ref{sec:potential} discusses the capabilities of \euclid{} for exoplanet microlensing and transit studies. In Section~\ref{sec:preimage} we discuss how early \euclid{} imaging of the \rom{ GBTDS fields would greatly enhance the precision of mass measurements for many \rom{} exoplanets. In Section~\ref{sec:simul} we discuss the huge benefits to \rom{} exoplanet science that can come from simultaneous observations of the GBTDS by \euclid{} and \rom{}. Section~\ref{sec:issues} discusses logistical issues and the potential impacts for \rom{} mission planning. Finally, in Section~\ref{sec:sum} we summarize the main points of the White Paper and list recommendations for consideration.

\section{The ESA Euclid telecope} \label{sec:telescope}

\euclid{} comprises a 1.2~m aperture Korsch-design telescope, equipped with two instruments. The optical \vis{} camera comprises 36 4096$\times$4132 CCD arrays with 0.1~arcsec pixels and a field of view of $0.79 \times 0.72$~deg$^2$ \citep{2016SPIE.9904E..0QC}. \vis{} has a single broadband filter spanning $0.55-0.9~\mu$m, approximately corresponding to $r+i+z$ filter bands, and will deliver a mean image resolution of 0.23~arcsec. The Near-Infrared Spectrometer and Photometer (\nisp{}) will undertake broad band near-IR imaging in $Y$ ($0.95-1.21~\mu$m), $J$ ($1.17-1.57~\mu$m) and $H$ ($1.52-2.00~\mu$m) filters \citep{2022A&A...662A..92E}, as well as spectroscopy over $0.92-1.85~\mu$m with a spectral resolution $R \simeq 250$. The \nisp{} imager comprises 16 $2048\times 2048$ arrays with a 0.3~arcsec pixel size and a $0.76\times 0.72$~deg$^2$ field of view. \euclid{} is designed for a nominal 6-year operation but may well be able to continue science activities significantly beyond this period. 

Both \euclid{} and \rom{} are scheduled to be injected into Earth-Sun L2 halo orbits, with a potential separation between them of up to 600,000~km. The actual separation between the spacecraft will depend on the precise separation of their launch dates and on the cumulative effect of any corrections made to their trajectories towards L2, and therefore cannot be precisely controlled. For the science discussed in this White Paper, we assume a separation in excess of 100,000~km \citep{2022A&A...664A.136B}.

\section{Direct planet mass measurements from space-based microlensing} \label{sec:micro}

\begin{figure}
\includegraphics[width=\textwidth]{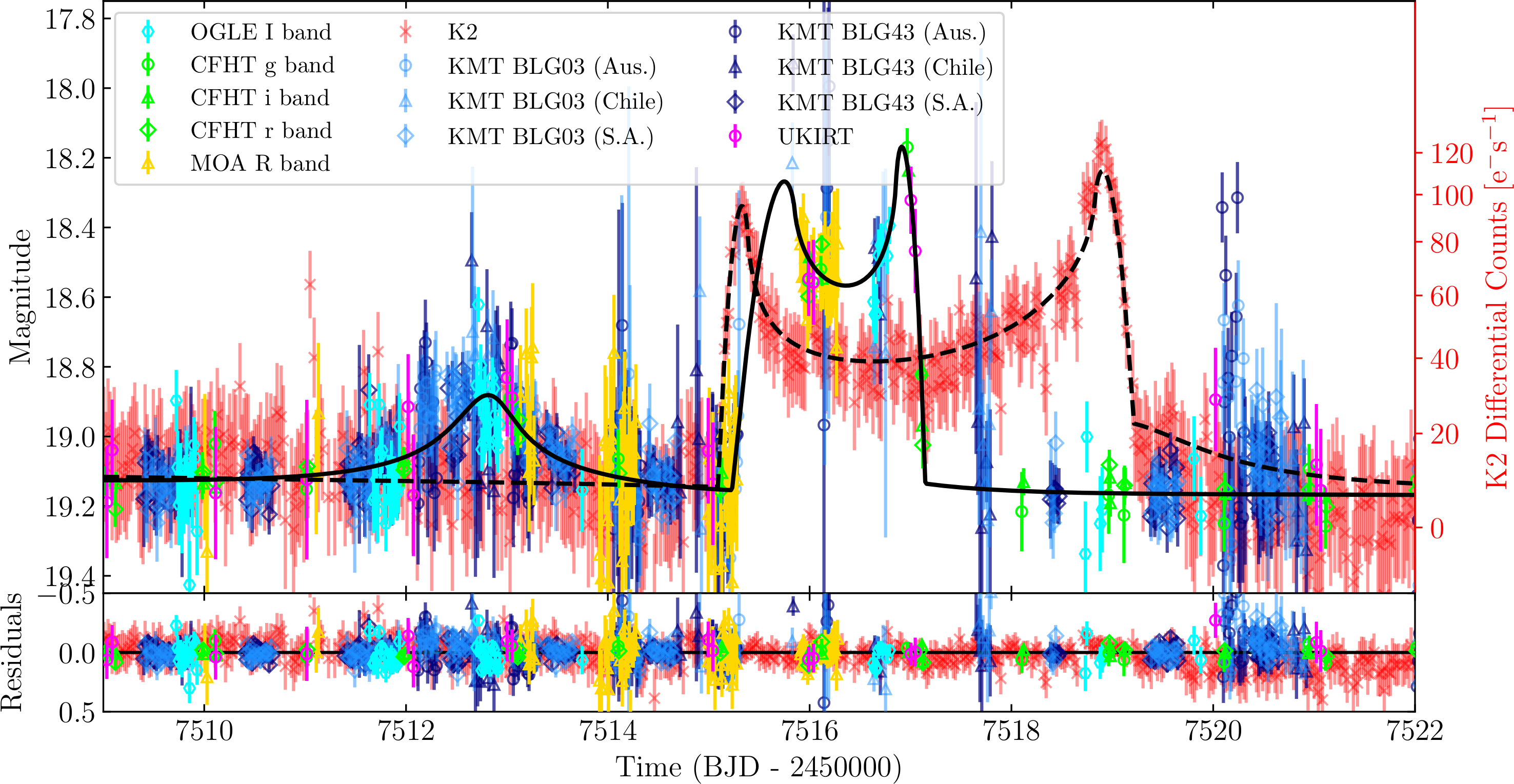}
\centering
\caption{The \kepler{} K2 Campaign~9 discovery lightcurve of K2-2016-BLG-0005Lb, the first bound exoplanet discovered from space using microlensing \citep{2023MNRAS.520.6350S}. The dense temporal sampling of the \kepler{} data (in red) was crucial for measuring the lens--source relative proper motion. The lightcurve was combined with data from a host of ground based observatories (coloured) in order to measure the lens--source relative parallax. These, along with the measured Einstein radius crossing time and high-resolution photometry of the source, enabled the host and planet masses, as well as the system distance, to be measured with high precision. This planet is one of the closest known analogues to Jupiter in terms of its mass and orbital radius, orbiting a $0.6~M_{\odot}$ host at a distance of 5.2~kpc. This is a surprising discovery, since massive planets orbiting low-mass stars are expected to be rare under core accretion theories of planet formation \citep{2021A&A...656A..72B}.  The exquisite flux sensitivity, temporal sampling and spatial resolution of the \rom{} GBTDS should enable direct planet mass measurements to be made for a large fraction of events. However, coordinated observations from \euclid{} will significantly boost sensitivity to proper motions and can provide simultaneous parallax measurements that will be crucial for confirming and robustly characterizing the tentative Earth-mass FFP population, whose presence is hinted in \kepler{} and current ground-based data \citep{2017Natur.548..183M,2021MNRAS.505.5584M}.}
\label{figs:K2C9-boundPlanet}
\end{figure}
The power of \rom{} for exoplanet microlensing lies not only in its ability to detect large numbers of exoplanets, but that, for a large fraction of these, it will be able to obtain a direct mass measurement of the planet. This is something that is extremely difficult to achieve from the ground, due to the effects of seeing and stellar crowding. It is this combination of increased sample size and the measurement precision quality of each event that will make the \rom{} sample so powerful for exoplanet demography \citep{2019ApJS..241....3P}.

There are a number of routes to obtaining a planet mass measurement from microlensing observations, and \rom{} is designed to be able to exploit them all at some level. The fundamental observable for all microlensing events is the event duration, given by the Einstein radius crossing time:
\begin{equation}
     t_{\rm E} = \frac{\theta_{\rm E}}{\mu_{\rm rel}},
     \label{eq:te}
 \end{equation}
where $\theta_{\rm E}$ is the angular Einstein radius and $\mu_{\rm rel}$ is the lens--source relative proper motion. $\theta_{\rm E}$ is given by \citep{2000ApJ...542..785G}
 \begin{equation}
     \theta_{\rm E} = \sqrt{\kappa M \pi_{\rm rel}},~~~\kappa = \frac{4G}{c^2 \mbox{au}}, ~~~
     \pi_{\rm rel} = \biggl(\frac{D_{\rm L}}{\mbox{au}}\biggr)^{-1}-\biggl(\frac{D_{\rm S}}{\mbox{au}}\biggr)^{-1},
     \label{eq:thete}
\end{equation}
where $M$ is the lens mass, $D_{\rm L}$ and $D_{\rm S}$ are the distances from observer to lens and source, respectively, $G$ is the gravitational constant and $c$ is the speed of light in vacuum. 
Equations~\ref{eq:te} and \ref{eq:thete} indicate that, to determine $M$, it is necessary to be able to measure three quantities from the four that appear in Equations~\ref{eq:te} and \ref{eq:thete}: $t_{\rm E}$; $\pi_{\rm rel}$; $\mu_{\rm rel}$; and $\theta_{\rm E}$. 

For a planetary microlensing system, $t_{\rm E}$ is governed not by the lensing action of the planet but by the duration of the lensing effect of the host, and is typically measured in weeks, or even months. It can always be obtained from a fit to the light curve for any adequately sampled event. 

Since the source star is visible, it is often possible to obtain some estimate for $D_{\rm S}$. This means that to determine $\pi_{\rm rel}$ we need to determine the lens distance $D_{\rm L}$. This can be achieved in some cases with deep high-resolution imaging long before or after an event where the intrinsic flux of the planetary host may be observable. Similarly, for visible lenses, and with excellent astrometric precision that is achievable from space, it may be possible to determine $\mu_{\rm rel}$ from images taken some years before or after an event is observed. Hence, direct mass measurements may be possible whenever the lens flux centroid can be determined with sufficient precision.

In cases where the lens and source fluxes are not separable, the lens mass may still be obtainable using subtle features that may be present on the lightcurve itself. In the case of planetary microlensing, magnification caustics may be used to resolve the angular size of the background source star through finite source size effects. This measurement provides the source star size in units of $\theta_{\rm E}$ and can be included as an additional fit parameter to the lightcurve. The absolute angular size of the source can be determined through angular size versus surface brightness colour relations \citep[e.g.][]{2004A&A...426..297K}. Therefore, comparison of the two quantities yields $\theta_{\rm E}$. 

Lastly, two well-separated observatories can be used to obtain simultaneous lightcurve measurements through different locations within the lens Einstein radius, allowing $\pi_{\rm rel}$ to be determined through triangulation. This is because the observer-lens-source geometry will differ, and so the two observatories will view lightcurves with different timing and magnification. 

Despite not being designed for observations of crowded stellar fields, a bound microlensing exoplanet, K2-2016-BLG-0005Lb, was recently uncovered and characterized from \kepler{} photometry \citep[][- see Figure~\ref{figs:K2C9-boundPlanet}]{2023MNRAS.520.6350S}, together with four lightcurves consistent with previously unknown Earth-mass free-floating planet (FFP) candidates \citep{2021MNRAS.505.5584M}. In the case of K2-2016-BLG-0005Lb, telescopes on the ground observed features that were visible earlier, and which were more condensed in time, than those seen by \kepler{}, due to observed differences in the trajectory of the background source across the planet--host binary lens system. Despite its very poor spatial resolution for crowded field studies (over 1400 Roman WFI pixels are needed to cover the area of 1 \kepler{} pixel), this discovery by \kepler{ validates the} underlying approach of space-based microlensing searches by showing how precision planet mass measurements can be obtained from stable crowded-field photometry and dense, uninterrupted temporal coverage that is achievable only from space. 

\section{Euclid's potential for microlensing and transit searches} \label{sec:potential}

The \euclid{} primary science mission will focus on the study of dark energy. However, possibilities for additional science have been explored in detail within the \euclid{} Consortium from its inception, and were included within the Science Definition Study \cite[the so-called ``Red Book'' -- ][]{2011arXiv1110.3193L}. A major difference between \euclid{} and \rom{} is that exoplanetary microlensing forms a core science activity for \rom{} that contributes to driving its design requirements; for \euclid{}, only the main cosmology mission can drive design requirements. However, the requirements for the \euclid{} weak lensing science ensure that \euclid{} is also very capable for microlensing. Possibly the biggest design constraint is the solar aspect angle, which limits the duration of continuous observations towards locations near the Ecliptic to around 23~days for nominal survey observations, and no more than twice yearly. This is significantly less than the $2 \times 72$-day constraint that applies to \rom{}, though it is still sufficient to adequately measure the duration of the bulk of microlensing events.

\begin{figure}
\includegraphics[width=0.75\textwidth]{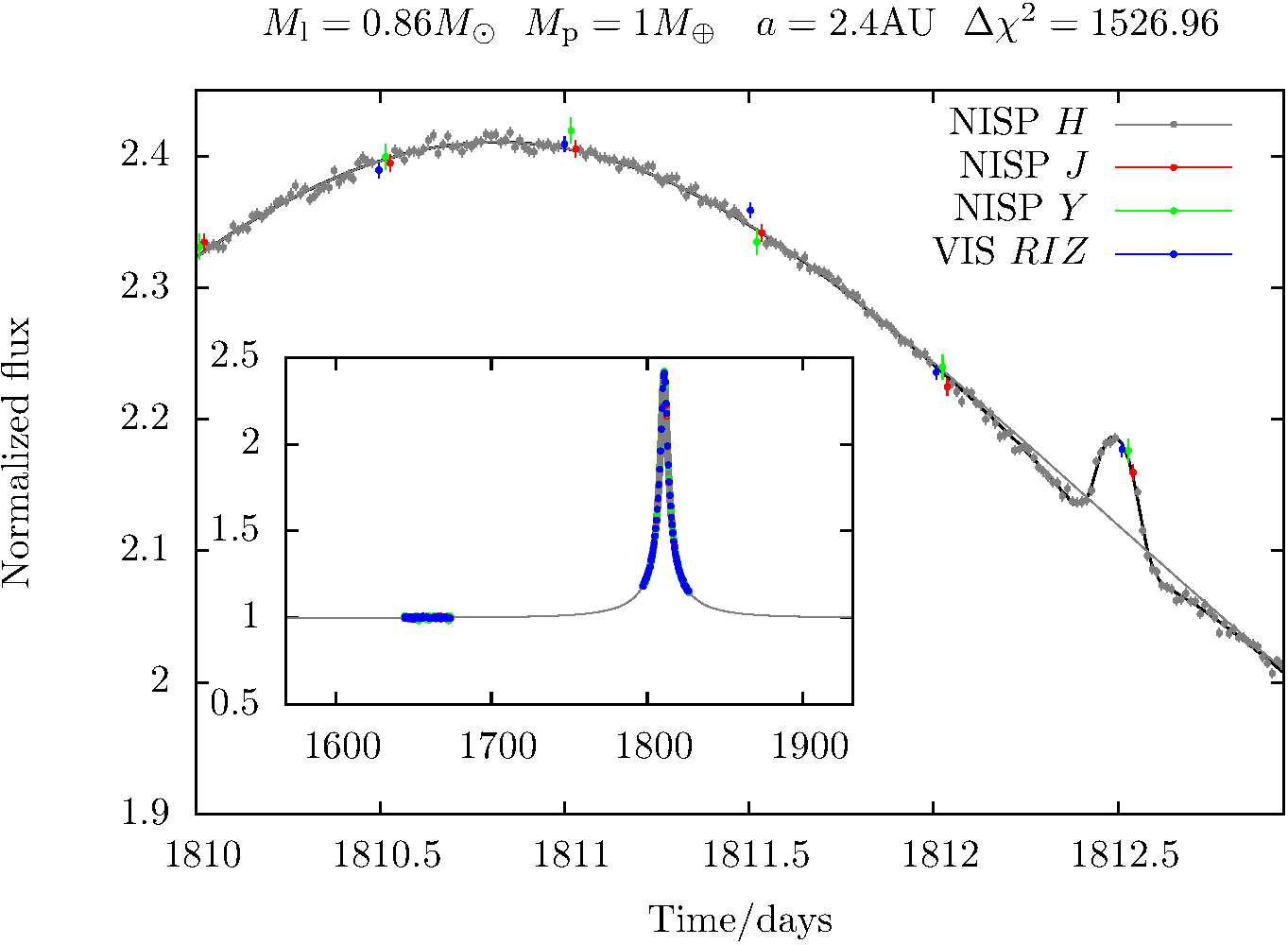}
\centering
\caption{A simulated planetary microlensing event as seen by \euclid{} in \nisp{}-$H$ with sparse sampling in \nisp{} $J$, $Y$ and \vis{} $riz$. The perturbation in the lightcurve is due to a $1~M_{\oplus}$ planet orbiting at 2.4~au around an early K-type/late G-type dwarf. Whilst low-mass cool planets are currently not accessible to other detection techniques, their microlensing perturbations can be easily detected by \rom{} and \euclid{} \citep[figure from][]{2013MNRAS.434....2P}.}
\label{fig:SimEvent}
\end{figure}

The \euclid{} Exoplanets Science Working Group has undertaken detailed design studies of reference surveys \citep{2013MNRAS.434....2P,2014MNRAS.445.4137M} that illustrate the capabilities of \euclid{} for exoplanet and transit discoveries. Figure~\ref{fig:SimEvent} shows an example of the exquisite microlensing lightcurve photometry that \euclid{} can deliver.
More recently, and as discussed in Sections~\ref{sec:preimage} and \ref{sec:simul}, we have shown how \euclid{} and \rom{} could work together to enhance the overall exoplanet science \citep{2022A&A...664A.136B}. 

It is our view that the most effective use of \euclid{} for microlensing observations would come from a powerful coordinated campaign with \rom{}. We envisage this to be possible in two phases: a pre-imaging survey of the \rom{} field at an early stage of the \euclid{} mission; and, as part of an extended \euclid{ mission}, a sustained campaign of simultaneous observations with \rom{} during some of the GBTDS observing seasons.
We discuss both of these phases below.

\section{Euclid pre-imaging of the Roman GBTDS area} \label{sec:preimage}

\begin{figure}
\includegraphics[width=0.495\textwidth]{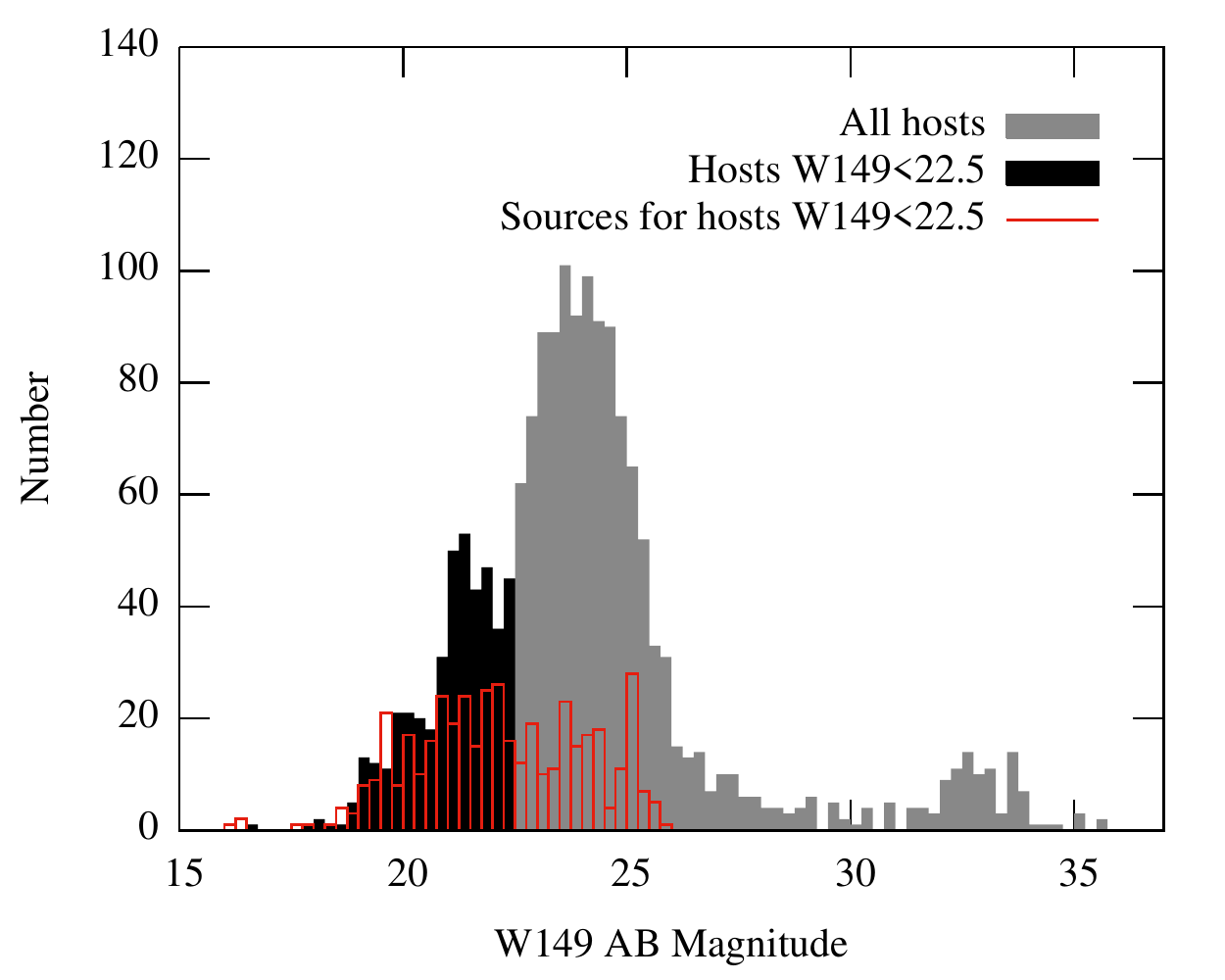}
\includegraphics[width=0.495\textwidth]{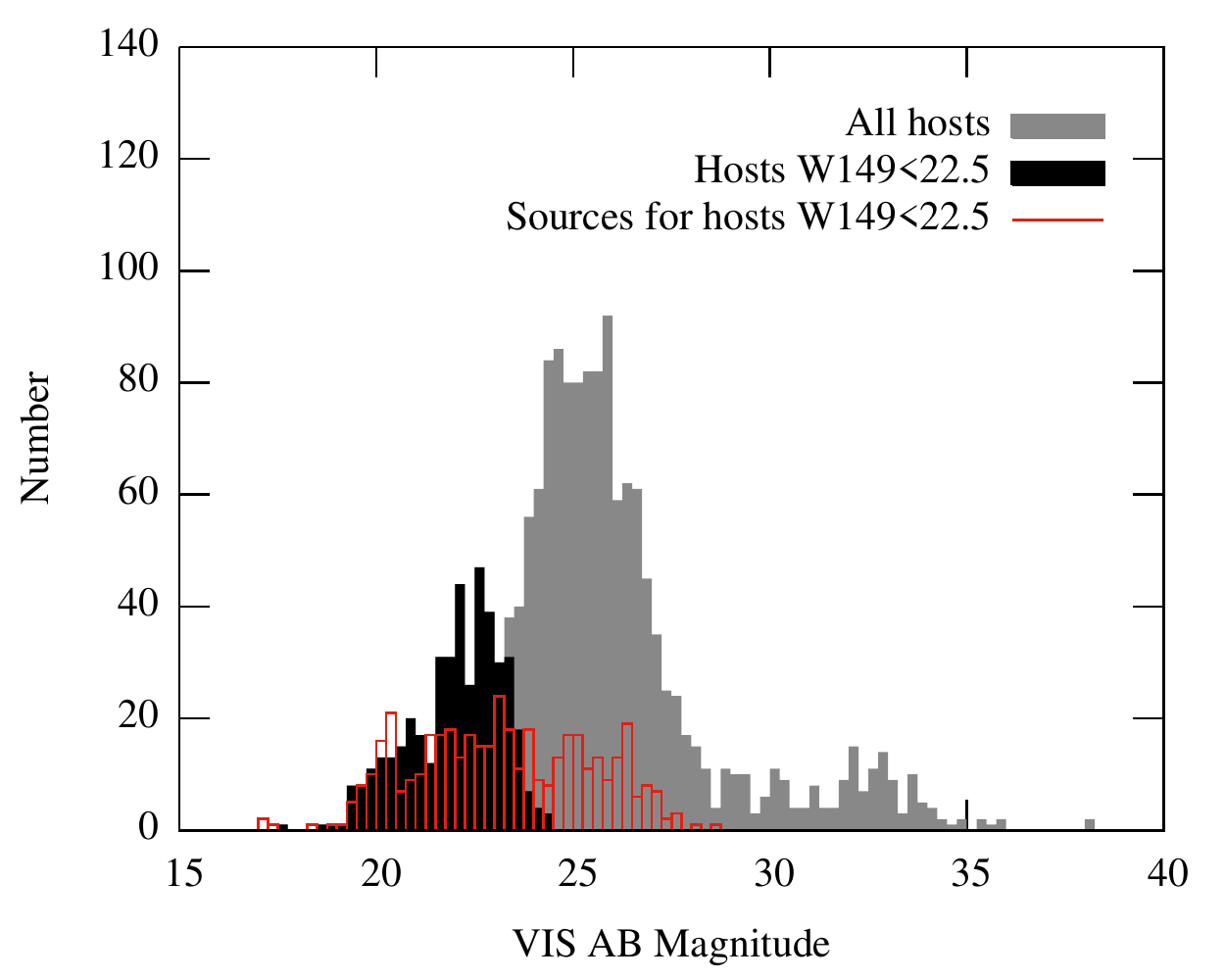}
\centering
\caption{Magnitude distributions of the lensing host stars extracted from simulated planetary microlensing events. Gray
histograms show \rom{} WFI~F146 filter magnitudes (left) and \euclid{} \vis{} (right) magnitudes for all lensing hosts within a sample of 1691 planetary microlensing events drawn from
simulations presented in \cite{2019ApJS..241....3P}, with the black histogram shows the brightest 432 events with F146~$ < 22.5$. Red histograms show the baseline magnitude distribution of the magnified background source stars to these 432 events. \citep[Figure from][]{2022A&A...664A.136B}}
\label{fig:EucRom-mags}
\end{figure}

\begin{figure}
\includegraphics[width=\textwidth]{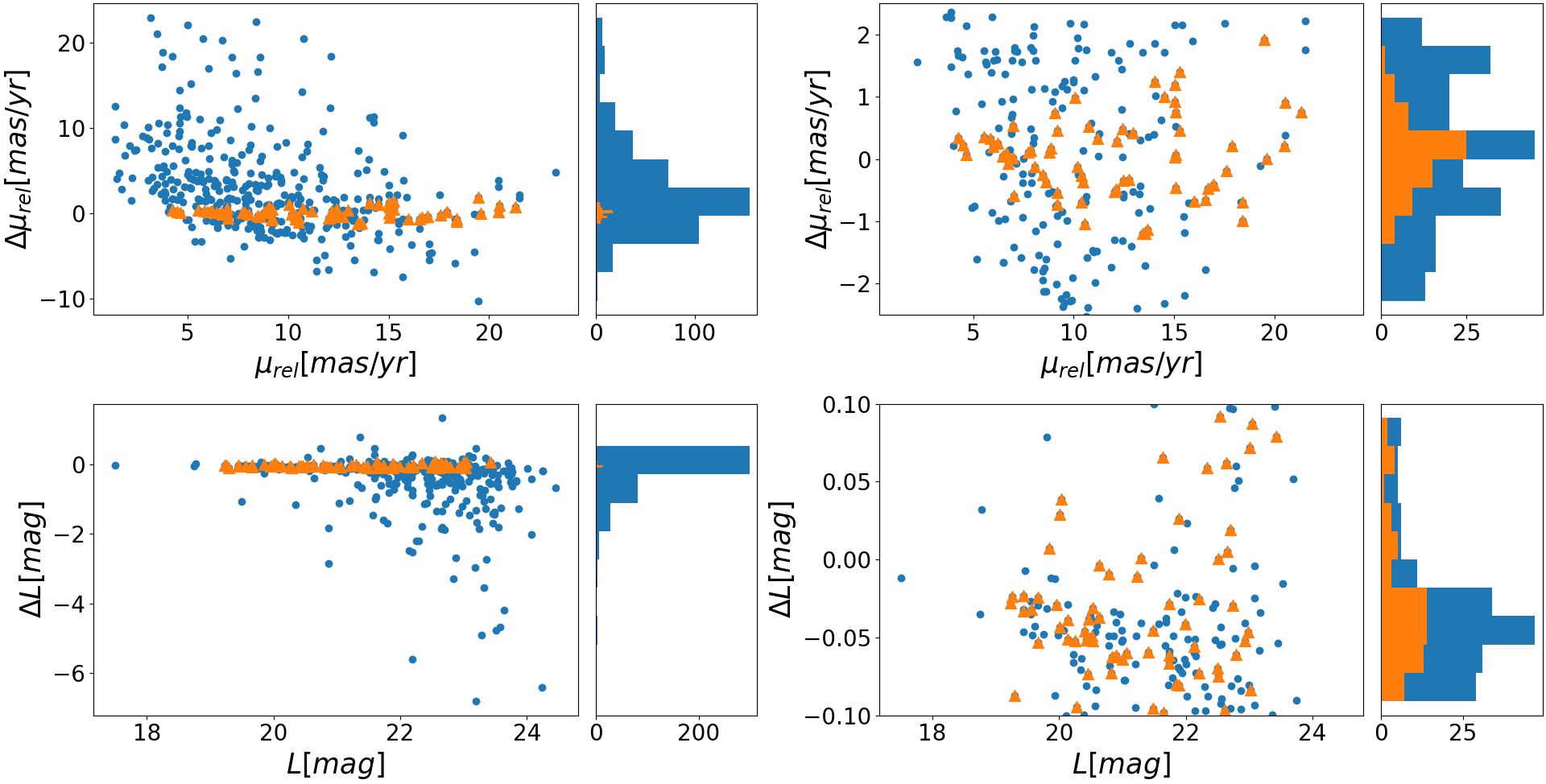}
\centering
\caption{The relative precision with which joint observations can measure the lens--source relative proper motions $\Delta \murel / \murel$ and lensing host magnitude $\Delta L / L$, determined from simulated photometry. The left column represents the results for all the 432 events with \rom{} F146~$< 22.5$, with orange points showing where the relative precision for both quantities is below $10\%$. The right column is a zoom for events with $|\Delta \murel | <2.5$ mas/yr and $|\Delta L|<0.1$ mag. More than 50 out of the 429 events satisfy both criteria and are well resolved, allowing precision mass measurements to be made. \citep[Figure from][]{2022A&A...664A.136B}}
\label{figs:EucRom-resids}
\end{figure}

\euclid{} will launch at least 3.5 years before \rom{} and therefore could provide early pre-imaging of the \rom{} fields. As discussed in Section~\ref{sec:micro}, this would enable lens fluxes to imaged by \euclid{} prior to microlensing events being subsequently observed by \rom{}. This would increase the precision of both the lens flux and $\mu_{\rm rel}$ measurements obtained from the astrometric motion of visible host lenses.

We have proposed that \euclid{} conducts a short single-epoch imaging survey of the \rom{} GBTDS area \citep{2022A&A...664A.136B}. Whilst the exact GBTDS area is still something to be defined, the microlensing science requirements of \rom{} mean that the general area is known. The \euclid{} mini-survey would comprise a series of dithered exposures covering the GBTDS general footprint in \vis{} and possibly with \nisp{}. This concept is under evaluation by ESA but could be executed during the \euclid{} calibration phase that would begin a month or so after launch.

In \cite{2022A&A...664A.136B} we envisage a survey involving 16 dithers of each of 4 \euclid{} pointings that would be needed to cover most or all of the GBTDS region. Careful stacking of the dithered images would provide an effective pixel size of $\sim 0.03$~arcsec. For our reference survey, we considered a 300~sec exposure time for \vis{}, with no \nisp{} observations. This would permit a full imaging sequence of the GBTDS area to take only around 7~hours.

The current Reference Observation Sequence (ROS) for \euclid{} imaging is fixed to a cycle that involves a 570~sec \vis{} exposure followed by a sequence of $J$, $H$ and $Y$ exposures of 112~sec duration each \citep{2022A&A...662A.112E}. The ROS involves a slew and settle period, followed by four iterations of this sequence that are dithered. The total time for these 4 dithers to be fully executed is expected to be 4214~secs. If we are constrained to this cycle for the mini-survey, and allowing for an average of 182~sec for a slew and settle between pointings \citep{2022A&A...662A.112E}, the time to execute the mini-survey would be about 19~hours in total.  Even with such a small commitment of time, the scientific returns are huge for exoplanet microlensing studies.

The \vis{} camera will offer the highest resolution (Section~\ref{sec:telescope}) and so potentially be most useful for astrometric precision. However, \nisp{} may also be useful as a better wavelength match to the near-IR WFI F146 filter that will be used for the GBTDS. Nonetheless, Figure~\ref{fig:EucRom-mags} shows that there is a very good overlap between sources observable to \rom{} in the F146 filter and the same sources as seen by the \euclid{} \vis{} camera. This gives some flexibility in how \euclid{} could conduct a mini survey. For instance, if persistence issues were a potential problem for \nisp{ observing the bulge region, then observations by the \vis{} instrument alone would be sufficient. This could be achieved without a significant alteration by selecting the close position in the filter wheel assembly during the \nisp{} YJH sequence.

\rom{} observations themselves will be obtained over a 5-year period, and so astrometric measurements of \rom{} data alone can be used to measure lens fluxes and proper motions for many events. However, the importance of additional early imaging observations by \euclid{} is two-fold. 

Firstly, the precision of the proper motion and lens flux measurements scale as the inverse cube of the proper motion baseline. This is because it rests on an accurate measurement of the skewness of the strongly overlapping lens and source point spread functions \citep{2007ApJ...660..781B}. For a \euclid{} mini-survey conducted 3.5 years prior to the start of a 5-year \rom{} campaign, this implies an improvement in the precision of these quantities by a factor $(8.5/5)^3 \simeq 5$ over what \rom{} can achieve alone. A corollary of this is that the advantage diminishes as the cube of the separation in time between the mini-survey and the launch of \rom{}. Figure~\ref{figs:EucRom-resids} shows the results of our simulation of joint \euclid{}--\rom{} lens flux and proper motion recovery \citep{2022A&A...664A.136B}.

Secondly, early pre-imaging will enable a much more rapid science return from the \rom{} dataset. Whilst \rom{} data will be publicly released as it is processed, reliable direct mass determinations from astrometric measurements would not be available from \rom{} data alone until several years into the \rom{} GBTDS campaign. With early \euclid{} imaging, it will be possible for some early mass measurement results to be obtained within the first year. By the end of the GBTDS, these and subsequent astrometric measurement results will be determined with much higher precision because of \euclid{} pre-imaging. 

\section{A Roman--Euclid Simultaneous observing campaign} \label{sec:simul}

Towards the end of the main cosmology mission, or immediately afterwards as an extended mission, \euclid{} could undertake simultaneous observations with \rom{} of the GBTDS area. We briefly summarize below several areas in which such observations would enable breakthrough discoveries. 

\subsection{Precision demography free-floating planets}

\begin{figure}
\includegraphics[width=\textwidth]{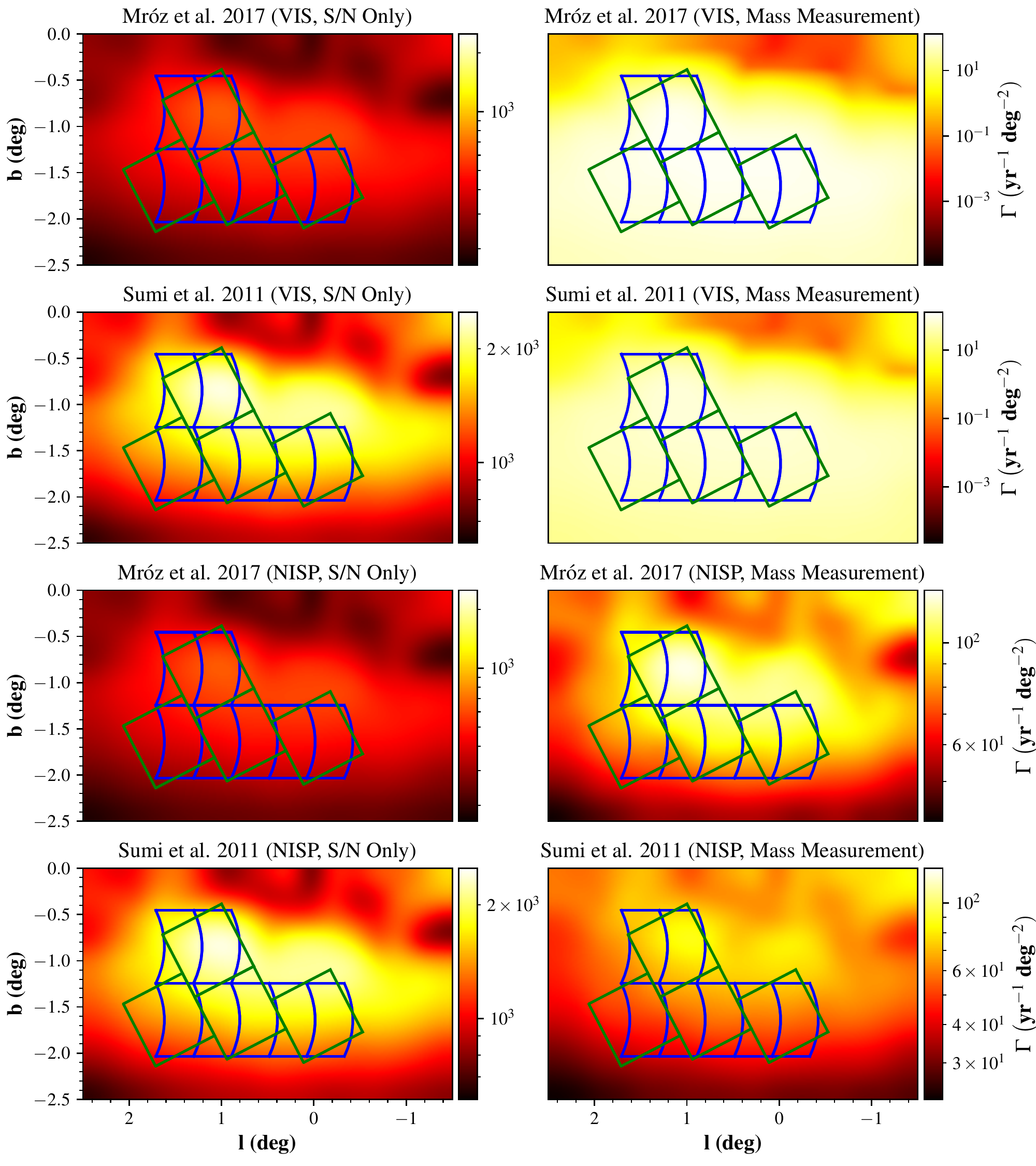}
\centering
\caption{The expected rate of FFPs detectable jointly by \euclid{} and \rom{}. Seven \rom{} pointings are shown in blue, following the survey design in \cite{2019ApJS..241....3P}. Four \euclid{} \vis{} and \nisp{} pointings will be able to cover most or all of this region, depending on relative orientation. Left panel shows the rate of detections for all FFPs above the simulation signal-to-noise detection threshold, whist the right column shows the rate of event of sufficient quality to obtain a direct mass measurement \citep[from][]{2022A&A...664A.136B}}
\label{fig:EucRom-rate}
\end{figure}

\begin{figure}
\includegraphics[width=\textwidth]{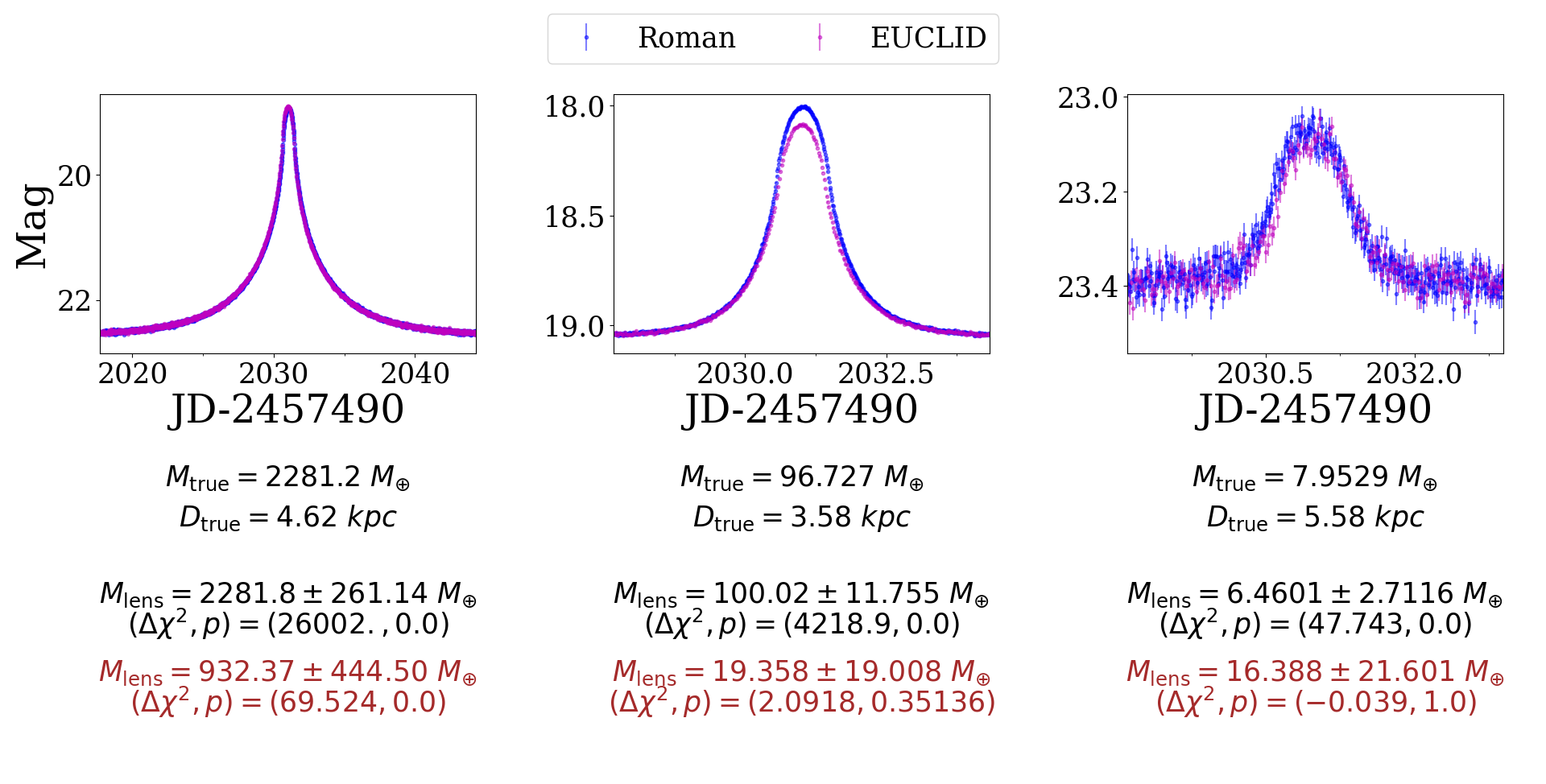}
\centering
\caption{Simulated photometry illustrating the unparalleled exquisite photometric capabilities of both \rom{} and \euclid{} for microlensing studies of FFPs. The separation of the two observatories in the L2 halo orbit allows for small parallax effects to be seen in the timing and maximum magnification of the lensing signals. They also allow high precision measurement of the finite-source effect -- the effect of differential magnification across the face of the background magnified source that causes a noticeable flattening of the event peak. The combination of parallax and finite source measurements, together with the Einstein radius crossing time, provide a direct mass measurement and therefore an unambiguous identification of FFP candidates Underneath each lightcurve the first row provides the true FFP mass, the next line is the joint fit to \euclid{} and \rom{} photometry and the lest row is the fit to \rom{} photometry alone. In each case, there is a very large systematic offset (factor 3-5) between fitted and true value when only \rom{} data is utilized \citep[from][]{2022A&A...664A.136B}. \euclid{} simultaneous observations will be crucial for robust demographic studies of FFPs.}
\label{figs:EucRom-FFP}
\end{figure}

Free-floating planets (FFPs) are planet-mass objects that are not bound to a host star. Candidates have been observed within nearby star forming regions \citep{2017ApJ...842...65Z}, whilst \kepler{} and ground-based microlensing data also suggest the existence of a potentially large Galactic population of FFPs. However, the ground based surveys have provided contradictory results. Early data from the MOA survey \citep{2011Natur.473..349S} indicated an excess of microlensing events with timescales of around 1~day, characteristic of Jupiter-mass lenses. The frequency in their dataset indicated a large population of two Jupiter-mass FFPs per Galactic star. More recently, an analysis of a larger dataset by OGLE \citep{2017Natur.548..183M} found no evidence of an excess at 1~day but did see tentative evidence for an excess of events of a few hours duration. Their excess, if true, would be consistent with a population of perhaps five Earth mass FFPs per Galactic star. Additionally, the OGLE and KMTNet survey teams have also obtained high time resolution data on individual events that are consistent with Earth-mass FFPs \citep{2020ApJ...903L..11M}, and such signals are also seen in an analysis of the \kepler{} K2 Campaign 9 microlensing survey \citep{2021MNRAS.505.5584M}. 

\rom{} alone can only measure the timescales of free-floating planet microlensing events, which provide an order of magnitude estimate of the planet’s mass, and in some cases the angular Einstein ring radius, though at present its 6- or 12-hour colour cadence is too slow to accurately and confidently measure $\theta_{\rm E}$ in many cases of events with microlensing timescales and source crossing timescales of <~few hours (e.g., as little as 10\% for Mars-mass free-floating planets, \citealt{2020AJ....160..123J}). \euclid{} \vis{} and \nisp{} observations can provide instantaneous colour measurements at high cadence ($\sim 30$~min) to guarantee $\theta_{\rm E}$ measurements in events with finite source effects. The separation of the two spacecraft in wide orbits around L2 is also well suited to measure the microlens parallax of low-mass lenses \citep{2019ApJ...880L..32B,2019BAAS...51c.563P,2020MNRAS.494.3235B,2022A&A...664A.136B}. Figure~\ref{fig:EucRom-rate} shows the expected joint detection rate of FFPs by \euclid{} and \rom{}, whilst Figure~\ref{figs:EucRom-FFP} provides examples of simulated joint lightcurves. 

Whilst simultaneous ground-based observations can be useful, they are no substitute for \euclid{} observations because of their shallow depth relative to \rom{} and their poor duty cycle at the times that \rom{} will observe the GBTDS ($\sim 3-6$~hours/night).
 
\subsection{Confirmation and mass measurement of exomoons} \label{sec:exomoon}

\begin{figure}
\includegraphics[width=\textwidth]{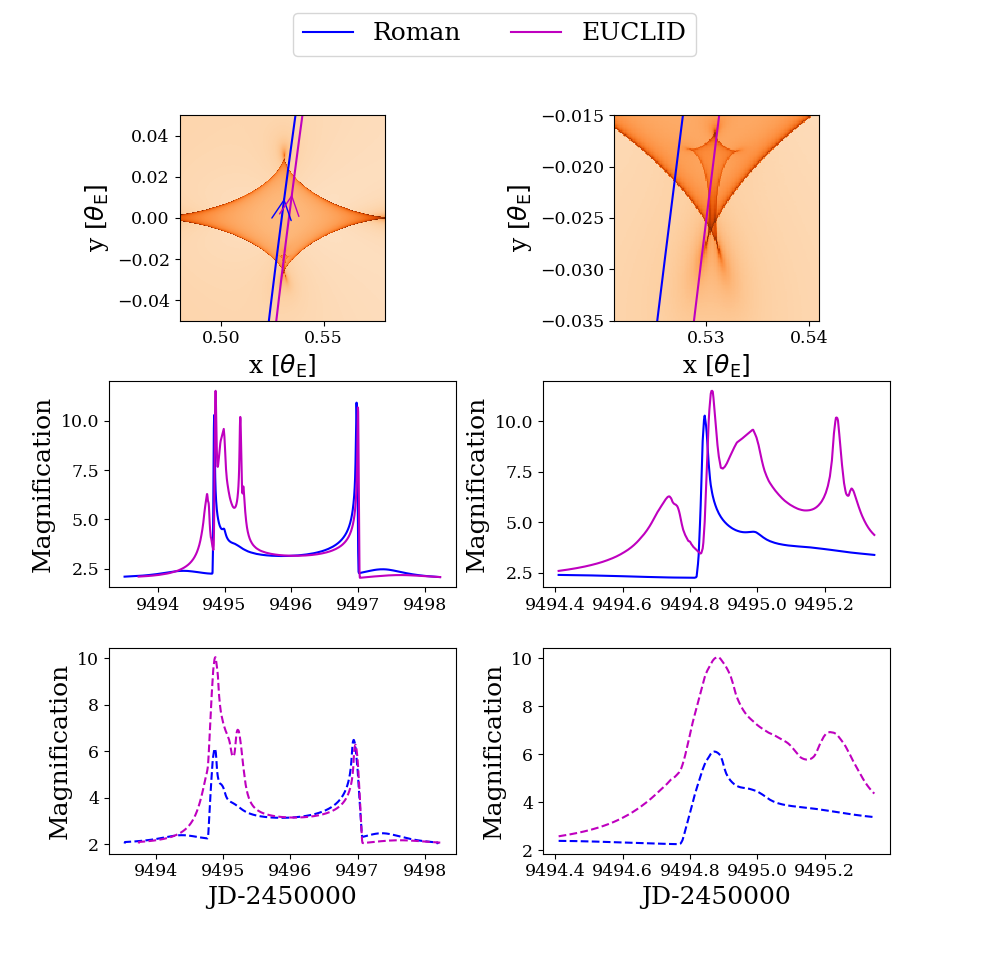}
\centering
\caption{Examples of theoretical lightcurves expected for microlensing planets that also exhibit anomalies due to the presence of an exomoon. The addition of an exomoon results in a 3-body lens system that can produce highly complex lightcurves. Simultaneous parallax observations can diverge sharply, enabling the modeling of such complex systems. Joint observations by \euclid{} and \rom{} would enable ground-breaking studies of such systems \citep[from][]{2022A&A...664A.136B}.}
\label{fig:EucRom-exomoons}
\end{figure}

\rom{} has the capability to detect the short-lived and typically low-amplitude signals of exomoons in microlensing events, but confirming an exomoon signal will be challenging with \rom{} data alone. The small number of photometric measurements over the signal and the large number of additional parameters that an exomoon adds to the model (at least three, the exomoon mass ratio, exomoon-exoplanet projected separation and angle, and potentially more to account for the exomoon’s orbital motion) mean that finding a unique exomoon model may be difficult. Additionally, a poorly sampled exomoon signal can be approximated by other scenarios, such as multiple planets or multiple source stars. Simultaneous \euclid{} observations provide high-cadence colour measurements that can rule out multiple source star scenarios, and the \rom{}-\euclid{} parallax provides a slightly different apparent path of the source through the exomoon-produced magnification pattern (Figure~\ref{fig:EucRom-exomoons)}. This enables disambiguation between exomoon vs multiple exoplanet scenarios and provides a measurement of the microlens parallax and angular Einstein ring radius, and thus mass of the exomoon. Again, simultaneous ground-based observations can be helpful, but the random nature of such signals within crowded fields often require the resolution and sustained temporal sampling only possible from space. Simultaneous observations from L2 by \euclid{} and \rom{} would be a game changer.
 
\subsection{Parallax and mass measurements for Roman’s bound planets} \label{sec:bound}

The \rom{} GBTDS is designed to enable mass measurements for most of its planet hosts by observing the lens and source separate over the mission’s 5-year baseline. This technique faces a number of challenges, however. A lens-source relative proper motion of 5~mas yr$^{-1}$ results in a separation after 5 years of just 0.25 of a \rom{} WFI pixel. This small shift makes measurements impossible if the lens is too faint, or if  companion stars to the lens or source bias measurements of the lens magnitude or the relative proper motion. It is essential therefore to maximize the number of events for which independent measurements can either independently yield the planet and host mass, or provide checks on masses derived through proper motion measurements. The separation of \euclid{} from \rom{} at L2 will enable parallax measurements for essentially any planetary signal that both telescopes observe simultaneously \citep{2019ApJ...880L..32B,2022A&A...664A.136B}, and roughly half of these will have angular Einstein ring radius measurements from finite source effects, so simultaneous \euclid{} observations can provide a valuable sample of independent mass measurements for \rom{}.
 
\subsection{False-positive rejection for Roman transiting planets} \label{sec:transits}

The GBTDS will detect 60,000-200,000 distant transiting exoplanets \cite{2023arXiv230516204W}. However, transiting planets can be mimicked by eclipsing binary stars with an M dwarf component of a similar size to a Jupiter-sized planet, or by faint eclipsing binary stars blended with a brighter star (a common occurrence in the GBTDS crowded fields). In both cases, the false positive signals produce chromatic primary eclipses and secondary eclipses. The \rom{} GBTDS will only make a small fraction of its measurements in other filters, so some fraction of its transiting planet candidates may not have enough colour measurements to distinguish false positives. High-cadence, simultaneous \euclid{} \vis{} observations can provide sufficient colour measurements for any transits that occur within the \euclid{}--\rom{} common survey area.

\section{Logistical constraints and potential impacts for Roman mission planning} \label{sec:issues}

The unique and high-value science provided by simultaneous observations with \rom{} GBTDS means that every opportunity to simultaneously observe with both telescopes should be taken. With their current scheduled launch dates and mission durations (\euclid{}: 07/2023--06/2029, \rom{} $\sim$10/2026--03/2032) there will be 2-3 years of main mission overlap, plus any \euclid{} mission extension. In the final years of its mission, the \euclid{} main survey schedule becomes inefficient \citep{2018SPIE10707E..12G} and it may be possible to schedule 1 or more microlensing seasons with \euclid{} that may be for the full duration allowed by its solar aspect angle and thermal constraints, or for a somewhat shorter duration to fit on with the ongoing main survey. In an extended \euclid{} mission, it may be possible to relax the \euclid{} pointing constraints to enable longer microlensing seasons. Exoplanet microlensing is one of the \euclid{} legacy science areas, but the scheduling of any such observations will depend on the progress and success of the \euclid{} main mission, so the exact timing can not be predicted at this time. Accommodating them may therefore require GBTDS season scheduling flexibility for \rom{}.
 
We anticipate two scenarios that could impact GBTDS scheduling, and one that could impact GBTDS observing strategy:
\begin{itemize}
\item {\bf Shifting whole GBTDS seasons within the main mission.} The primary goals of GBTDS dictate maximizing the separation between its first and last seasons within the nominal mission duration, but the seasons between these are more flexible, though should still be placed as close to the start or end of the mission as possible. The high-latitude time domain survey (HLTDS) is also likely to have a contiguousness constraint of 2 years of continuous observations every 5 days. Ideally, GBTDS and HLTDS observations should not intersect, but for simultaneous \euclid{}-\rom{} observations, the disruption to a GBTDS season could be a price worth paying. Quantitatively, if HLTDS requires 1 day out of 5 for observations and overhead, this would cause a $\sim 20\%$ decrease in the GBTDS planet yield in each season that was affected by interleaved observations.
\item {\bf Addition of GBTDS seasons.} If \euclid{} is granted an extended mission, it could provide the opportunity for $\sim 6$ seasons with simultaneous \euclid{}-\rom{} observations of the GBTDS during the \rom{} main mission. Somewhere between 2 and 4 of these are likely to be regularly scheduled GBTDS seasons, but the others would need to be added, potentially exceeding the initial allocation to the GBTDS. We will not comment on the feasibility of this, but can suggest a way in which the necessary added time is minimized. For the free-floating planet case, only \rom{} observations simultaneous with the \euclid{} shorter ($\sim 20-30$~day) observing window would be necessary. For the bound planet and exomoon science cases, it would be necessary to gather observations over the \rom{} full 72 day observing window in order to characterize the host microlensing event; however, outside the simultaneous \euclid{} observations, this could be conducted at a much-reduced cadence, e.g., as infrequently as once per week in order to trace out the longer host star microlensing event.
\item {\bf Adjusting the GBTDS fields and cadence to maximize Roman-Euclid parallax measurements.} The detector layouts of \euclid{} are not ideally matched to the likely geometry of the GBTDS fields. Additionally, \euclid{} has a limited set of observing modes, which provide long visit times and subsequently longer cadence. It may make sense to observe a smaller number of GBTDS fields during simultaneous \euclid{} observations in order to maximize the number of small planets and exomoon detections observed by both spacecraft. Detailed quantification of this is challenging, though potential methods are available  \citep{2019ApJS..241....3P} that could be used to assess the optimum number of fields to observe to maximize the detection rate of planets of various masses.
\end{itemize}

All of the described modifications would require the \euclid{} mission to commit to major observational campaign(s) during or after the nominal mission, and it is too early to say whether this will be possible.

\section{Summary and recommendations} \label{sec:sum}

Through early pre-imaging of the \rom{} GBTDS fields, and by conducting some simultaneous microlensing observation near the end are after the \euclid{} primary cosmolgy science, \euclid{} observations can greatly enhance the exoplanet science that \rom{} will deliver. 

We have outlined the following capabilities that would be enabled by \euclid{} coordinated observations:
\begin{itemize}
    \item Through a 19-hour imaging survey of the \rom{} GBTDS fields undertaken in the first few months of the launch of \euclid{}, mass measurements for many  \rom{} exoplanet microlensing events could gain an improvement in precision by up to a factor of $\sim 5$.
    \item An extended \euclid{} mission would enable \euclid{} to undertake observations of the GBTDS simultaneously with \rom{}. This would: 
    \begin{itemize}
        \item Enable \rom{} and \euclid{} to make direct mass measurements of free-floating planets and a robust demographic survey of the Earth-mass free-floating planet regime hinted at by Kepler and ground-based microlensing surveys;
        \item Create a vastly improved capability for \rom{} to detect {\em and confirm} exomoons, and to measure their mass;
        \item provide an enhanced capability for \rom{} and \euclid{} to make direct planet mass measurements through space-based simultaneous parallax observations;
        \item Enable greater false-positive discrimination of transiting planets.
    \end{itemize}
\end{itemize}

We wish to emphasize to both NASA and ESA communities the incredible opportunity for the study of exoplanet demography presented by the imminent launch of \euclid{}, and by the possibility of simultaneous observations during the overlap lifetime of \euclid{} and \rom{}. We believe that it may be beneficial for NASA and ESA to explore these opportunities through a formal Joint Study Group.

An important consideration is that a sustained microlensing campaign with \euclid{} will not be possible until near, or after, the end of the cosmology primary science mission, which means it would like not happen until at least five years after launch (i.e. late summer 2028). By this stage, \rom{} is scheduled to be near the end of its second year of  operations, and so may already have completed three out of the six scheduled GBTDS seasons. This would leave a maximum of three seasons of coordinated observation (one season in late 2028 and the others from 2031, if \euclid{} is still operating). Each of these can be a maximum of 30 days due to the \euclid{} observational limits on the solar aspect angle. 

An alternative possibility would be to consider a flexibility in the sequencing of the GBTDS seasons in order to maximize the number of coordinated \euclid{}--\rom{} observing seasons. However, such a move may only make sense if ESA is able to formally agree to commit the required time to a sustained \euclid{} microlensing campaign.

Given the technical and logistical uncertainties, at this stage the most important requests of this White Paper are:
\begin{enumerate}
    \item For NASA and ESA to convene a \rom{}-\euclid{} Joint Study Group that can evaluate how to maximize the science potential of the two telescopes, within the operating and scheduling constraints of both;
    \item for \rom{} to maintain the flexibility to reschedule GBTDS seasons during the mission;
    \item for the Committee to consider a mechanism and possible cost to GBTDS to add seasons to take advantage of opportunities for \euclid{}-\rom{} simultaneous observations (this could be via GO allocations, but given the necessary interagency coordination, we expect this could be challenging to implement).
\end{enumerate}

\bibliography{main}


%

\end{document}